# Comparison of thermal fluctuations in foam films and bilayer structures

Nikoleta G. Ivanova and Roumen Tsekov
Department of Physical Chemistry, University of Sofia, 1164 Sofia, Bulgaria

In the frames of the DLVO theory the root mean square amplitude and correlation length of capillary waves in thin liquid films are calculated. Their dependencies on some important physical parameters are studied. Two models are considered: films with classical interfaces and films between lipid bilayers. The performed numerical analysis demonstrates essential difference in their behavior, which is due to the different elastic properties of the film surfaces in the models.

Thin liquid films (TLFs) bounded by simple interfaces, insoluble monolayers and membranes are in the scientific focus for many years [1]. The interest to these nano-thin structures is promoted by specific phenomena taking place in TLFs and their large application in technology. An important aspect in the practical application is the stability of TLFs [2-4], which is decisive for many processes in flotation [5], colloidal coagulation [6], food industry [7] and other disperse systems [8]. In the literature, the TLF stability is described as a result of competition between the van der Waals disjoining pressure and stabilizing effect of capillary and electrostatic forces (DLVO theory). A linear hydrodynamic theory of the kinetics of hole or black spot formation is developed [9-11] which provides calculation of the film lifetime and critical thickness. This theory accounts for many interfacial processes such as adsorption, surface and bulk diffusion of surfactants, Marangoni effect, etc. The basic restriction of the linear theory is the small parameter expansion over the ratio $\zeta/h$ between the amplitude $\zeta$ of the surface corrugations and the average film thickness $h$. Since at the point of the film rupture they are of equal order such presumption is not justified. However, the linear theory results are very important for interpretation of light scattering experiments [12-15].

An interesting effect on the TLF stability is the coupling of the film drainage and the unstable modes due to non-linear TLF hydrodynamics [16, 17]. Its theoretical description requires correct calculation of the drainage rate, which depends substantially on the shape of the film interfaces. Usually, a dimple is formed [11, 18-21] and the corresponding law of thinning [22, 23] differs from the classical Reynolds expression applied first to TLF by Scheludko. During last decades, some attempts to take into account the influence of non-linear effects on TLF stability have been made. Some authors [24-26] have reported analytical-numerical calculations based on the Navier-Stokes equations. These theories continue usual ideas of hydrodynamic stability without considering fluctuations. Other authors [27] have accounted for the influence of the thermo-convection in a non-linear flow by means of temperature dependence of viscosity. The main non-linear effects are due to two factors: the non-linear hydrodynamic terms in the Navier-Stokes equations and the non-linear dependence of the disjoining pressure from the film thickness. In pure form, the first effect takes place at capillary-gravity waves on infinitely deep liquids where the disjoining pressure is missing. In this case, the interfacial kinematics is described by a Korteweg-de Vries equation of fifth degree [28]. The dependence of the disjoining pressure from the film thickness is a problem, which has been object of intensive investigations. Abreast of now becoming classical van der Waals and electrostatic components, a number of new interesting effects of macroscopic interaction were observed [29]. Here we can mention non-DLVO hydrophobic, hydration, protrusion and other forces [30]. In the case of surfactant concentration above CMC the disjoining pressure isotherm exhibits periodic behavior, which is responsible for the film stratification [31]. Similar periodic behavior but with smaller characteristic length could be expected as a consequence of the discrete molecular structure of the matter [32].

The picture in membrane multi-lamellar structures is additionally complicated by the low value of surface tension leading to large out of plane deviations of membranes. For this reason, the steric interaction between the two membranes bordering the film becomes important and leads to the so-called undulation forces [33]. This interaction depends substantially by the membrane elasticity $\kappa$, which can be conveniently measured by ellipsometry and X-ray scattering [34, 35]. Membranes are soft and easy flexible bilayer structures, which possess freedom for fluctuations into membrane plane [36]. The scale of these fluctuations grows up close to the intra-membrane phase transitions [37, 38]. The mechanical properties of membranes, even if composed by two or more monolayers of insoluble surfactants, are far from those of the latter [39, 40]. The internal behavior of bilayers is well studied by means of methods based on diffraction and scattering of light [41, 42], X-rays [43] and neutrons [44, 45], and some interesting phase transitions are observed as result of the phase state of the lyophobic parts of surfactant molecules [46, 47]. They play important role not only in the membrane transport, elastic properties and secondary quasi-crystalline arrangement of the membrane incorporations but also in the biological functions. Additional complications could arise from existence of membrane proteins, which interact each other by potential and curvature induced surface forces [21, 48, 49]. Due to the Brownian motion they migrate and create a secondary crystalline structure, which changes properties of the whole membrane [50]. The problem of rupture of TLF bounded by biomembranes is relevant to coalescence of biological cells, fusion of colloids in cells, stability of vesicles and lamellar structures and other life important processes. Both linear [51] and non-linear [52] theories were applied to the multi-lamellar membrane structure stability. A quasi-thermodynamic theory for rupture of bilayers and Newton black films was proposed [53], which explains the phenomenon by appearance of holes with a critical size.

The local fluctuations of the thickness of a thin liquid film are the major factor responsible for the film stability. Usually the thin liquid films behave rheologically as a Newtonian fluid and they can be described on the base of the classical hydrodynamics coupled with the fluctuation theory. In the frames of the well-known lubrication approximation the TLF hydrodynamics obeys the following Navier-Stocks-Reynolds equations

$$\partial_x p = \eta \partial_{zz} v_x \qquad \partial_y p = \eta \partial_{zz} v_y \qquad (1)$$

$$\partial_z p = 0 \tag{2}$$

$$\partial_x v_x + \partial_y v_y + \partial_z v_z = 0 \tag{3}$$

where the $z$-axis is normal on the film surface, $p$ is the hydrodynamic pressure including fluctuations, $v_x$, $v_y$ and $v_z$ are the components of the hydrodynamic velocity, and $\eta$ is the dynamic viscosity of the liquid in the film. Since the symmetric modes of local film thickness changes, the so-called squeezing modes [13], are only responsible for the film rupture, the lateral boundary conditions reads

$$v_x(z = \pm H/2) = 0 \qquad v_y(z = \pm H/2) = 0 \tag{4}$$

which imply films with tangentially immobile surfaces; $H(x, y, t)$ is the local film thickness. According to Eq. (2), the pressure does not change along the film normal, which allows a direct integration of Eqs. (1). The corresponding results are

$$2\eta v_x = (z^2 - H^2/4)\partial_x p \qquad 2\eta v_y = (z^2 - H^2/4)\partial_y p \tag{5}$$

which account for the boundary conditions (4). Introducing Eqs. (5) into the continuity equation (3) and integrating once under the obvious symmetry condition $v_z(z=0)=0$ yield an explicate expression for the normal component of the hydrodynamic velocity

$$6\eta v_z = \partial_x[z(3H^2/4 - z^2)\partial_x p] + \partial_y[z(3H^2/4 - z^2)\partial_y p] \tag{6}$$

Finally, combining the kinematic relations for the film surfaces

$$\partial_t H + v_x(z = \pm H/2)\partial_x H + v_y(z = \pm H/2)\partial_y H = v_z(z = H/2) - v_z(z = -H/2)$$

with Eqs. (5) and (6), the following equation governing the evolution of the local film thickness $H$ is derived

$$12\eta \partial_t H = \partial_x(H^3 \partial_x p) + \partial_y(H^3 \partial_y p) \tag{7}$$

In general, the local film thickness can be split to the mean value $h \equiv <H>$ and the local fluctuation $\zeta \equiv H - h$. In the frames of the considered non-draining flat films the mean thickness does not depend on the position at the film and is constant in time. A realistic model for such a film is the soap bubble. The rigorous theory of the stability of films requires solution of the nonlinear equation

(7). Unfortunately, the contemporary mathematics does not supply methods for a rigorous solution. For this reason all consideration which follows are limited in the frames of a linear model where $\zeta^2 \ll h^2$. Hence, the linearized version of Eq. (7) for the present films reads

$$12\eta \partial_t \zeta = h^3 \Delta p \tag{8}$$

Here a two-dimensional Laplace operator $\Delta \equiv \partial_{xx} + \partial_{yy}$ is introduced for convenience. Equation (8) is not complete and requires appropriate expression for the hydrodynamic pressure on the local film thickness. According to linear TLF thermodynamics the normal force balance on the film surface takes the form

$$p = -(\sigma/2)\Delta\zeta + (\kappa/2)\Delta\Delta\zeta - \Pi'\zeta + \delta P \tag{9}$$

where the reference pressure in the gas phase is taken to be equal to the disjoining pressure $\Pi(h)$ which is the necessary condition for stationary of the film average thickness. The first term on the right hand side of Eq. (9) represents the local capillary pressure, where $\sigma$ is surface tension. The second term accounts for elastic deformations of the film surfaces with elasticity modulus $\kappa$ of the interfaces. $\Pi'$ is the first derivative in respect to $h$ of the total disjoining pressure being a sum of all existing components, and $\delta P$ represents the pressure fluctuations due to the thermal motion of the matter.

By introducing Eq. (9) in Eq. (8) one yields the following basic equation

$$(24\eta/h^3)\partial_t \zeta + \sigma\Delta\Delta\zeta - \kappa\Delta\Delta\Delta\zeta + 2\Pi'\Delta\zeta = 2\Delta\delta P$$

describing the fluctuation dynamics of the surface waves. A useful tool for analysis of this stochastic equation is the Fourier transform in respect to the spatial coordinates. After its application the equation above changes to

$$(24\eta/q^2 h^3)\partial_t \zeta_q + (\sigma q^2 + \kappa q^4 - 2\Pi')\zeta_q = -2\delta P_q \tag{10}$$

where $q$ is the modulus of the wave vector and the new indicated quantities are the Fourier images of the original ones. Right now one can recognize the destabilizing role of the negative components of the disjoining pressure on the film waves. They can lead to a negative harmonic force in Eq. (10) which is an indication for divergent solutions. It is important to note the stabilizing effect of the surface tension and film surface elasticity.

From Eq. (10) it is possible to obtain the spatial spectral density of the thermal non-homogeneity of the film thickness which is defined via the relation [17]

$$C_{\zeta\zeta}(q,t) \equiv <\zeta_q \zeta_q^*> = \int_0^R K_{\zeta\zeta}(\rho R,t) J_0(q\rho R) 2\rho d\rho \qquad (11)$$

where $K_{\zeta\zeta}$ is the spatial autocorrelation of the waves, $R$ is the film radius and $J_0$ is Bessel function of first kind and zero order. The integral presentation in Eq. (11) employs the assumption for spatial uniformity of the film and for this reason the autocorrelation depends only on the distance between the two points of interest. Multiplying Eq. (10) by the complex conjugated Fourier component $\zeta_q^*$, taking average value of the result and using definition (11) one yields the following equation

$$(12\eta/q^2 h^3)\partial_t C_{\zeta\zeta} + (\sigma q^2 + \kappa q^4 - 2\Pi') C_{\zeta\zeta} = -2<\zeta_q^* \delta P_q> \qquad (12)$$

To close this equation an appropriate modeling of the last statistical model is required. It can be obtained from the classical theory of thermodynamic fluctuations. According to this theory the mean value of the product of any fluctuation quantity and its thermodynamically conjugated one (i.e. the partial derivative in respect to this quantity of the minimal work necessary to disturb the system equilibrium) is equal to $k_B T$, where $k_B$ is the Boltzmann constant and $T$ is temperature. In the present case the work of the fluctuation forces is equal to

$$\delta W = -\int \zeta \delta P dx dy = -\pi R^2 \sum_q \zeta_q^* \delta P_q \qquad (13)$$

where the last equality expresses a well-known theorem relating the integral sums of the originals and their Fourier images. One can conclude from Eq. (13) that the thermodynamically conjugated quantity of $\delta P_q$ is $-\pi R^2 \zeta_q^*$. Hence, according to the principle cited above, the following relation $<\zeta_q^* \delta P_q> = -k_B T/\pi R^2$ holds. Consequently, Eq. (12) acquires the form

$$(12\eta/q^2 h^3)\partial_t C_{\zeta\zeta} + (\sigma q^2 + \kappa q^4 - 2\Pi') C_{\zeta\zeta} = 2k_B T/\pi R^2 \qquad (14)$$

To obtain the solution of Eq. (14) it is necessary to know the initial condition. The assumption that initially the film has been in a state of infinitely deep liquid where the disjoining pressure vanish is plausible from the viewpoint of the film history. Hence, one can model the initial spectral density as one corresponding to an infinitely thick film at equilibrium

$$C_{\zeta\zeta}(q,0) = 2k_B T/\pi R^2 (\sigma q^2 + \kappa q^4) \qquad (15)$$

Thus, the solution of Eq. (14) with initial condition (15) reads

$$C_{\zeta\zeta}(q,t) = \frac{2k_B T}{\pi R^2 (\sigma q^2 + \kappa q^4)} \frac{2\Pi' \exp[(q^2 h^3 / 12\eta)(2\Pi' - \sigma q^2 - \kappa q^4)t] - \sigma q^2 - \kappa q^4}{2\Pi' - \sigma q^2 - \kappa q^4} \qquad (16)$$

This fundamental result contains the well-known quantitative criterion for the film stability. It is easy to recognize from Eq. (16) that the surface waves become unstable ($C_{\zeta\zeta}(q,t \to \infty) \to \infty$) if the inequality $2\Pi' > \sigma q^2 + \kappa q^4$ is fulfilled. The corresponding equality defines a critical wave number

$$q_{cr} = \sqrt{(\sqrt{\sigma^2 + 8\kappa\Pi'} - \sigma)/2\kappa}$$

below which the waves are not stable and their amplitudes grow exponentially. In the case of negligible elastic modulus $\kappa$ this criterion reduces to the well-known result $q_{cr} = \sqrt{2\Pi'/\sigma}$ derived by Scheludko first. The other extreme case of negligible surface tension is appropriate for lamellar membrane structures and the corresponding critical wave number is $q_{cr} = \sqrt[4]{2\Pi'/\kappa}$. These relations show again that unstable films are those with positive first derivative $\Pi' > 0$ of the disjoining pressure.

From Eq. (16) one can derive the value of the wave vector corresponding to the most rapid fluctuation mode, which is an important characteristic of the linear theory [11]. It follows from Eq. (16) the following expression for the decrement of the growth of the waves

$$\alpha \equiv \lim_{t \to \infty} \ln[C_{\zeta\zeta}(q,t)]/t = (2\Pi' - \sigma q^2 - \kappa q^4) q^2 h^3 / 12\eta$$

As seen, the antagonism between the destabilizing effect of the viscous friction and the stabilizing effect of the interfacial elasticity leads to maximum in the dependence of the decrement of wave growth on the wave vector. In this point the derivative of $\alpha$ becomes zero

$$\partial_q \alpha = (2\Pi' - 2\sigma q^2 - 3\kappa q^4) h^3 q / 6\eta = 0$$

The solution of this equation is the wave number of the most rapid Fourier mode

$$q_{mq} = \sqrt{(\sqrt{\sigma^2 + 6\kappa\Pi'} - \sigma)/3\kappa}$$

It is easy to show that the most rapid wave is unstable since $q_{mq} < q_{cr}$. For the two cases discussed before the corresponding values of the wave numbers of the most rapid waves are equal to $q_{mq} = \sqrt{\Pi'/\sigma}$ and $q_{mq} = \sqrt[4]{2\Pi'/3\kappa}$, respectively.

As was clearly demonstrated in the introduction, one of the most important actors in the dynamics of the TLF fluctuation waves is the disjoining pressure. Many publications have considered the van der Waals component as the main force leading to rupture of symmetric films. The most simple and relatively good expression for this component of the disjoining pressure is given by the following formula $\Pi_{VW} = -K_{VW}/h^3$, where $K_{VW}$ is a specific constant for the film matter. The negative value of $\Pi_{VW}$ corresponds to a tendency for film thinning, which determines the destabilizing effect of this force. The present paper will consider DLVO forces only as a first attempt to complete description of the fluctuation picture in the film. For this reason, an adequate expression for the electrostatic component of the disjoining pressure is required. For the sake of convenience, the following well-known formula $\Pi_{EL} = 64 k_B T N_A C \exp(-h/D)$ is employed, which is valid for relatively large surface potentials [1]. Here $N_A$ is the Avogadro number, $C$ is the concentration of a monovalent electrolyte in the film, and $D$ is the Debye length of the counter ion atmosphere. The latter is given by the expression $D^2 = \varepsilon_0 \varepsilon N_A k_B T / 2F^2 C$, where $\varepsilon_0$ and $\varepsilon$ are the dielectric permittivity of vacuum and matter of the film, and $F$ is the Faraday constant. After these assumptions the total disjoining pressure is a sum of the two components described above

$$\Pi = -K_{VW}/h^3 + 64 N_A k_B T C \exp(-\sqrt{2C/\varepsilon_0 \varepsilon N_A k_B T} \, Fh)$$

As was pointed put, in the frames of the linear theory the basic parameter accounting for the specific interactions in the film is the first derivative of the disjoining pressure in respect to the film thickness. For the present case it takes the form

$$\Pi' = 3K_{VW}/h^4 - 64CF\sqrt{2N_A k_B TC/\varepsilon_0 \varepsilon} \exp(-\sqrt{2C/\varepsilon_0 \varepsilon N_A k_B T} \, Fh) \qquad (17)$$

In the Figure 1 the isotherm of $\Pi'$ is presented for aqueous films with specific constants $K_{VW}$ = 6.25 meV, $\varepsilon_0 \varepsilon$ = 0.5 nF/m and $T$ = 300 K. Here, the film thickness is presented in nanometers and the concentration of electrolyte is in mol/l. It is clear from this figure that the increasing of the electrolyte lead to appearance of a positive derivative of the disjoining pressure. The latter corresponds to film instability, which is the explanation for the destabilizing role of electrolytes in the colloid coagulation. A more detailed analysis shows that there are two values of the concentration $C$ where the derivative $\Pi'$ is zero. It is well known that the lower one corresponds to a maximum of $\Pi$ while the upper value corresponds to a minimum of the disjoining pressure. The part of $\Pi$ being after the minimum does not have effect on the stability analysis of the film even if the derivative $\Pi'$ is positive. This fact is due of the appearance of the maximum at lower thickness, which in the frame of a nonlinear theory corresponds to a stable state. The real instability will be detected for film thickness smaller than that corresponding of the maximum. For the present case the transition between stable and unstable film takes place in the range of electrolyte concentrations lower than $C$ = 10 mM.

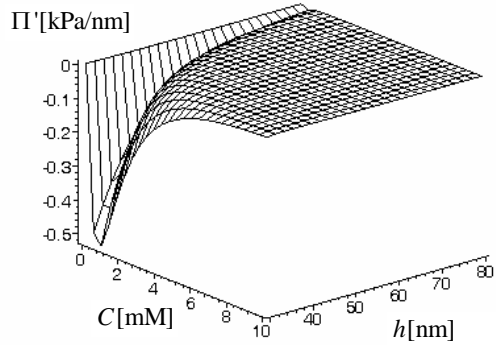

**Fig. 1** The dependence of $\Pi'$ on the electrolyte concentration and the film thickness

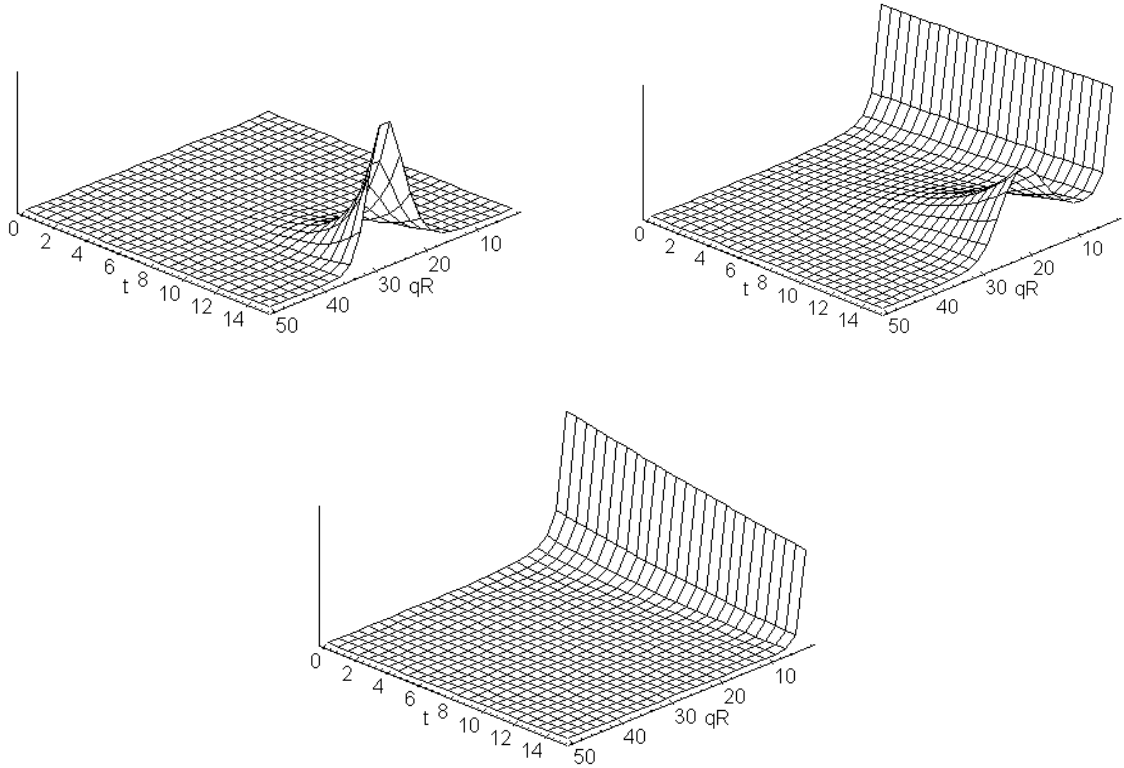

**Fig. 2** Evolution of the spectral density for model A at $C$ = 0, 2 and 10 μM

The spatial spectral density of the waves can be obtained from Eq. (16) completed by Eq. (17). The evolution of $C_{\zeta\zeta}(q,t)$ in time is presented below at different electrolyte concentrations. In the present paper two sets of specific constants are considered: model A for a film with fluid phase surfaces and model B for a film restricted by two membranes. In both cases the bulk fluid is water and for this reason the bulk specific and geometric characteristics of the films are equal: $T$ = 300 K; $\varepsilon_0\varepsilon$ =

0.5 nF/m; $K_{VW}$ = 6.25 meV; $\eta$ = 1 mPa s; $h$ = 30 nm; $R$ = 0.1 mm. The only difference is in the characteristic properties of the film interfaces. A typical value of the surface tension for the model A is $\sigma$ = 50 mN/m while the elastic modulus is negligible and for this reason $\kappa$ is accepted to be zero. In Figure 2 the evolution of the spectral density $C_{\zeta\zeta}(q,t)$ for the model A (the time is in seconds) as a function of the electrolyte concentration is represented. As seen, the film is unstable for electrolyte concentration 0.000 mM and its amplitude grows in time. In this case the electrostatic disjoining pressure is zero. At concentration 0.010 mM the film is stable and the spectral density of the waves corresponds to that for an infinite liquid. Films with concentration 0.002 mM are unstable but in the considered time interval the effects of van der Waals and electrostatic forces are commensurable. For this reason the resulting picture is a combination of the other two. From the figures above it is easy to conclude that the wave vector of the most rapid mode $q_{mq}$ is about 250 mm$^{-1}$, which corresponds well to the theoretical prescribed value of $q_{mq} = \sqrt{\Pi'/\sigma}$ = 272 mm$^{-1}$. The wave number of the critical wave $q_{cr} = \sqrt{2\Pi'/\sigma}$ = 385 mm$^{-1}$ is also underlined on the first two figures.

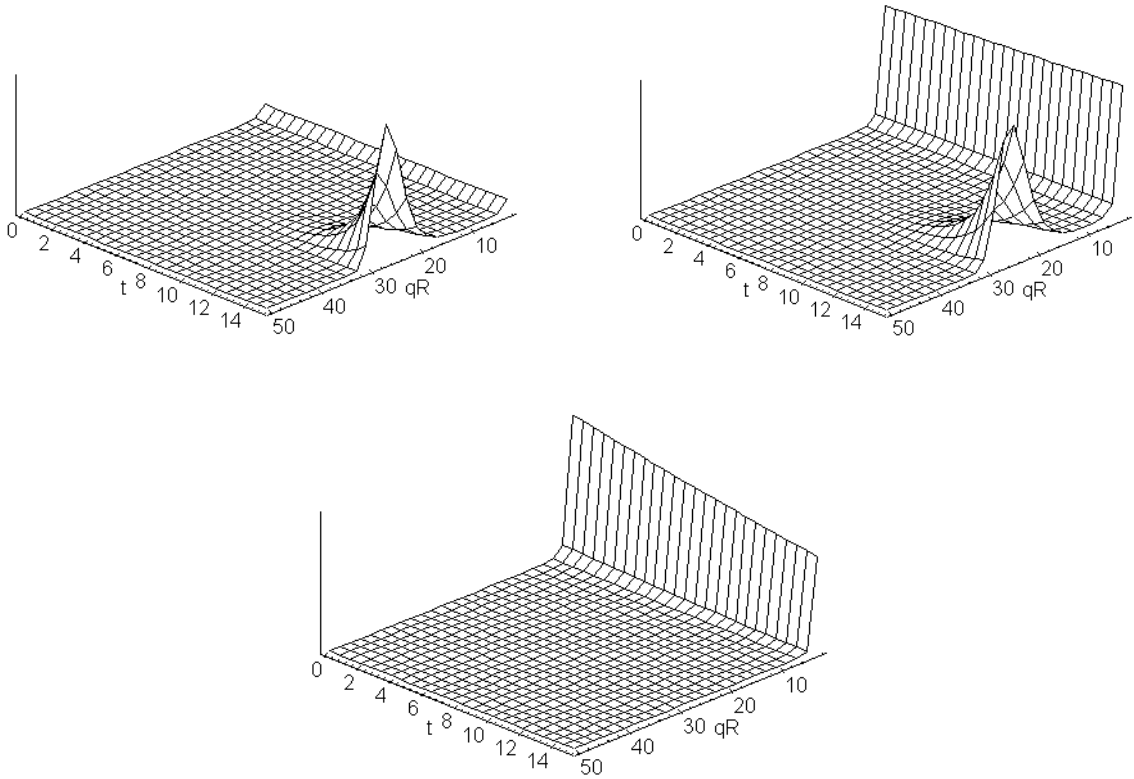

Fig. 3 Evolution of the spectral density for model B at $C$ = 0, 1 and 10 μM

The thin liquid films between lipid bilayers possess surface tension many times lower than the foam films but their interfacial elasticity modulus is large. A typical value of $\kappa$ is 0.5 pJ, while $\sigma$ can be considered to be zero. In Figure 3 the evolution of the spectral density for model B is presented for different electrolyte concentrations. The dependence of the spectral density on the electrolyte concentration is analogical to that of the model A. From the figures it is easy to determine the wave vectors of the most rapid and critical modes, which correspond well to the expectations $q_{mq} = \sqrt[4]{2\Pi'/3\kappa}$ = 265 mm$^{-1}$ and $q_{cr} = \sqrt[4]{2\Pi'/\kappa}$ = 345 mm$^{-1}$, respectively.

Knowing the spectral density one is able to calculate several important statistical characteristics of the fluctuation waves. For the stability analysis the most important are the root mean square amplitude of the waves $A$ and the number of uncorrelated subdomains $N$. They are defined via the relations [17]

$$A^2(t) \equiv <\zeta^2> = \sum_{i=1}^{\infty} C_{\zeta\zeta}(q=\lambda_i/R,t)/J_1^2(\lambda_i) \tag{18}$$

$$N(t) = A^4 / \sum_{i=1}^{\infty} C_{\zeta\zeta}^2(q=\lambda_i/R,t)/J_1^2(\lambda_i) \tag{19}$$

where $\{\lambda_i\}$ is the set of all zeros of the Bessel function of zero order, i.e. $J_0(\lambda_i)=0$, and $J_1(\cdot)$ is Bessel function of first class and first order. Since the analytical expressions for the sums above are not possible, they have to be calculated numerically. In the present paper only the first one hundred zeros $\{\lambda_i\}$ of the Bessel function are used in the calculations. A restriction of the number of zeros is allowed since the unstable modes are those with wave numbers smaller than $q_{cr}$. In all the cases here $\lambda_{100} \approx 313$ is at least twice larger than the critical value $q_{cr}R$.

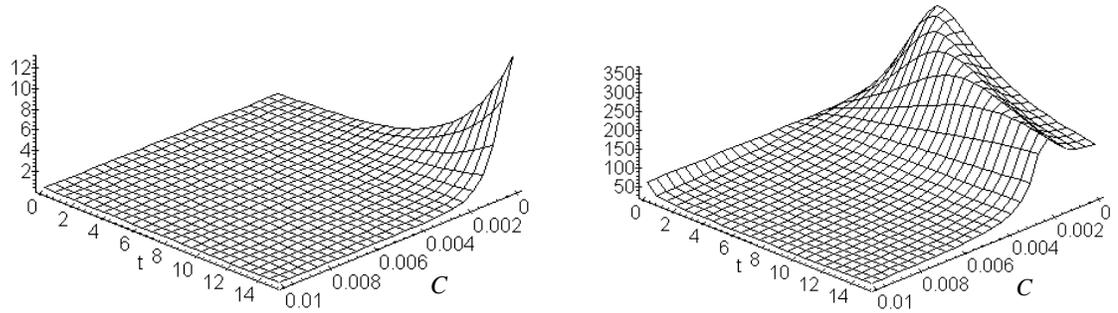

**Fig. 4** Evolution of the amplitude $A(t,C)$ in nm and number of domains $N(t,C)$ for model A

In Figure 4 the calculated mean square amplitude $A$ and number of uncorrelated domains $N$ for model A are presented as a function of the electrolyte concentration. The first figure demonstrates the exponential growth of the amplitude of the surface waves in time after the stable-unstable film phase transition with decreasing electrolyte concentration. More interesting is the second figure, which demonstrates breakdown of the film to many uncorrelated subdomains at the transition state. As was demonstrated in previous investigations [54, 55], the evolution of $N$ in time passes through a maximum, which is due to rearrangement of the wave spectrum. Analogical calculations for the model B are presented in Figure 5. The amplitude of the waves and the number of uncorrelated subdomains exhibit nearly analogical behavior like those of the model A. The basic difference is the relatively lower number of uncorrelated subdomains in model B. It is not surprising, however, since the wave vector dependence of the elastic term is on forth power, while the surface tension multiplies only by the second power of $q$. This leads to a sharper dependence of the spectral density on the wave vector at the model B, hence to a decrease of the number of the existing unstable modes. The latter reflects in a decreased number of uncorrelated subdomains $N$. It is important to note the relative retardation of the spectrum rearrangement at model B as compared to the model A.

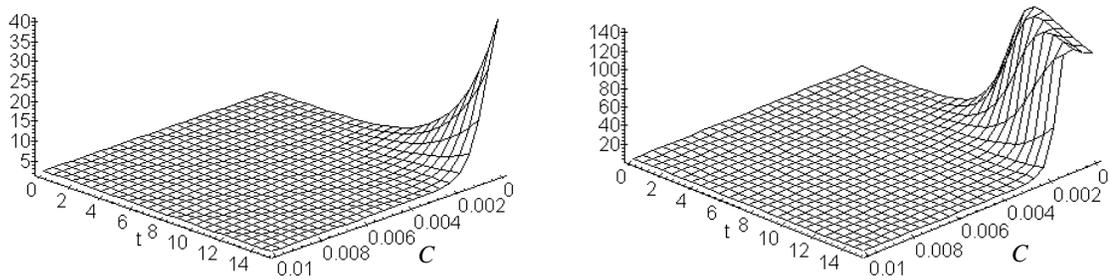

**Fig. 5** Evolution of the amplitude $A(t,C)$ in nm and number of domains $N(t,C)$ for model B

In order to discriminate the influence of different factors on the stability of TLFs, the dependences of $A$ and $N$ on the surface tension $\sigma$ (standard value 50 mN/m for model A), elasticity $\kappa$ (standard value 0.5 pJ for model B), film thickness $h$ (standard value 30 nm), viscosity $\eta$ (standard value 1 mPa s) and film radius $R$ (standard value 0.1 mm) will be considered for both models. The electrolyte concentration is fixed equal to $C$ = 1 mM. Let us start with the dependence on the surface tension. As expected the amplitude of the thickness waves increases strongly with decreasing surface tension. The latter is an important factor for stabilization of films and its decrease reflects in appearance of unstable waves and increase of the number of uncorrelated subdomains:

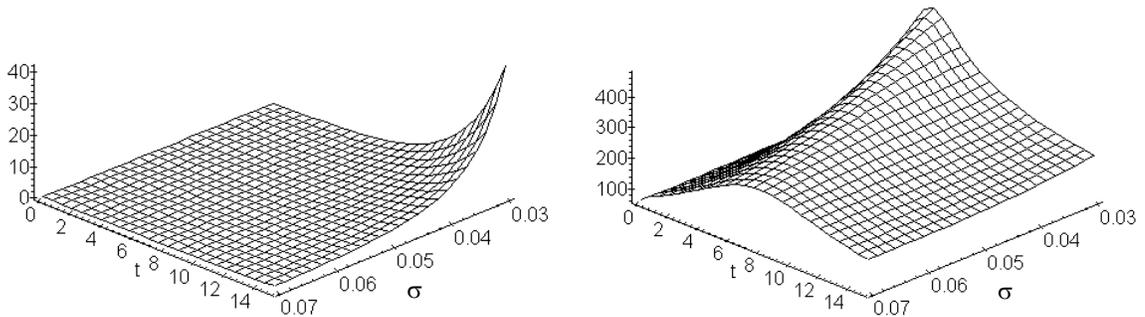

**Fig. 6** Evolution of the amplitude $A(t,\sigma)$ in nm and number of domains $N(t,\sigma)$ for model A

Let us see now the effect of the elasticity on the film stability. Calculations show that moderate changes in $\kappa$ do not affect the behavior of films for model A. This could be easily explained by the fact that the unstable modes correspond to lower values of the wave vector and for this reason the effect of the surface tension is emphasized. On Figure 7 the effect of $\kappa$ in pJ is studied on model B.

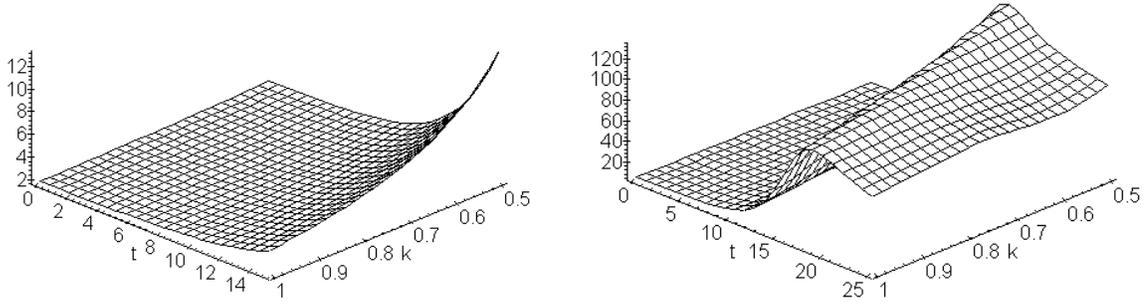

**Fig. 7** Evolution of the amplitude $A(t,\kappa)$ in nm and number of domains $N(t,\kappa)$ for model B

The first plot demonstrates the expected effect of suppression of the waves by increasing elasticity, analogical to the effect of $\sigma$. More interesting is the second plot, where the number of uncorrelated subdomains increases at smaller $\kappa$. The effect is, however, weaker than that of the surface tension in the model A, which can be attributed to the forth root in $q_{cr}(\kappa) = \sqrt[4]{2\Pi'/\kappa}$ as compared to the second root in $q_{cr}(\sigma) = \sqrt{2\Pi'/\sigma}$ for the model A.

Another important parameter for the dynamics of the film fluctuations is the mean film thickness, which is responsible for three effects. First, with decrease of $h$ the derivative $\Pi'_{VW}$ increases thus leading to film destabilization. Second, with decrease of $h$ the derivative $\Pi'_{EL}$ decreases, which is a stabilizing effect. And third, the decrease of $h$ strongly increases the friction coefficient of the fluctuations, which slows down their evolution. Because of the interplay of these effects it is expected an complicated dependences of $A$ and $N$ on $h$. Moreover, the decrement $\alpha$ could demonstrate affinity to some values of $h$.

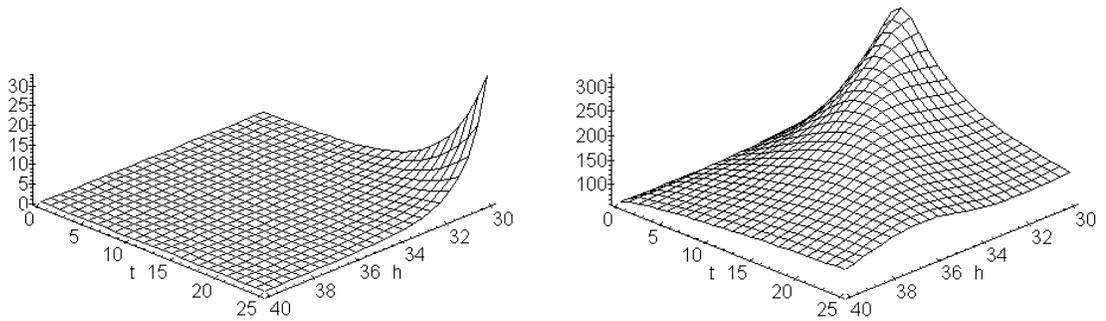

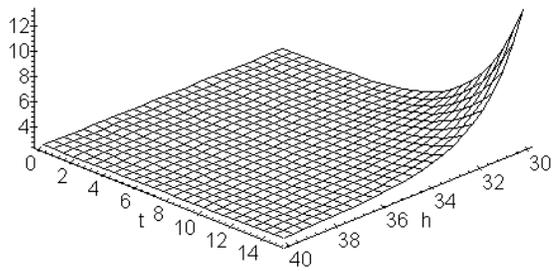 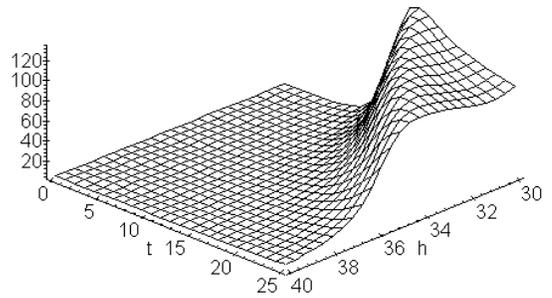

**Fig. 8** Evolution of the amplitude $A(t,h)$ in nm and number of domains $N(t,h)$ for models A and B

The plots of the root mean square amplitude clearly show the accelerated evolution of the unstable modes at thinner films. Moreover the number of uncorrelated subdomains grows also as a consequence of the increase of the value of the critical wave vector. Note that again the evolution of $N$ demonstrates a larger latent period for model B, after which it grows tremendously. According to Eq. (16) the role of the film viscosity is to slowdown the evolution of the waves. Since the values of the critical and the most rapid wave vectors do not depend on the viscosity, the latter should not affect strongly the number of uncorrelated subdomains, which is evident from the calculations below

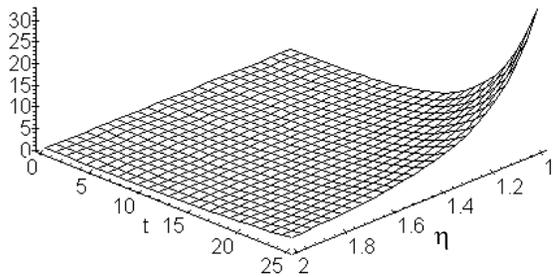 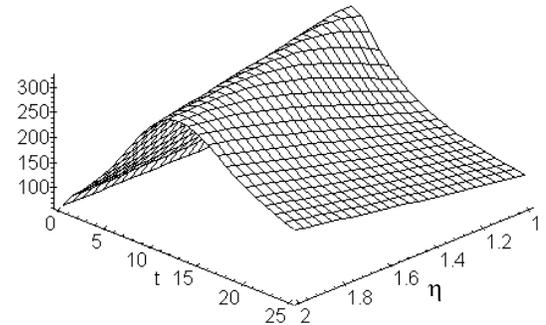

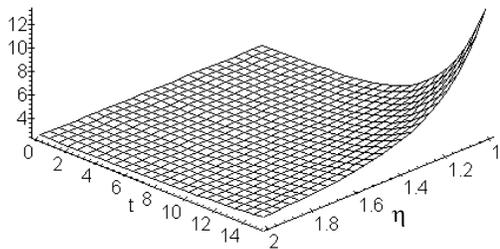 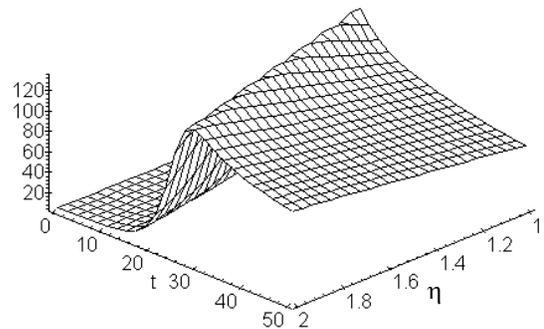

**Fig. 9** Evolution of the amplitude $A(t,\eta)$ in nm and number of domains $N(t,\eta)$ for models A and B

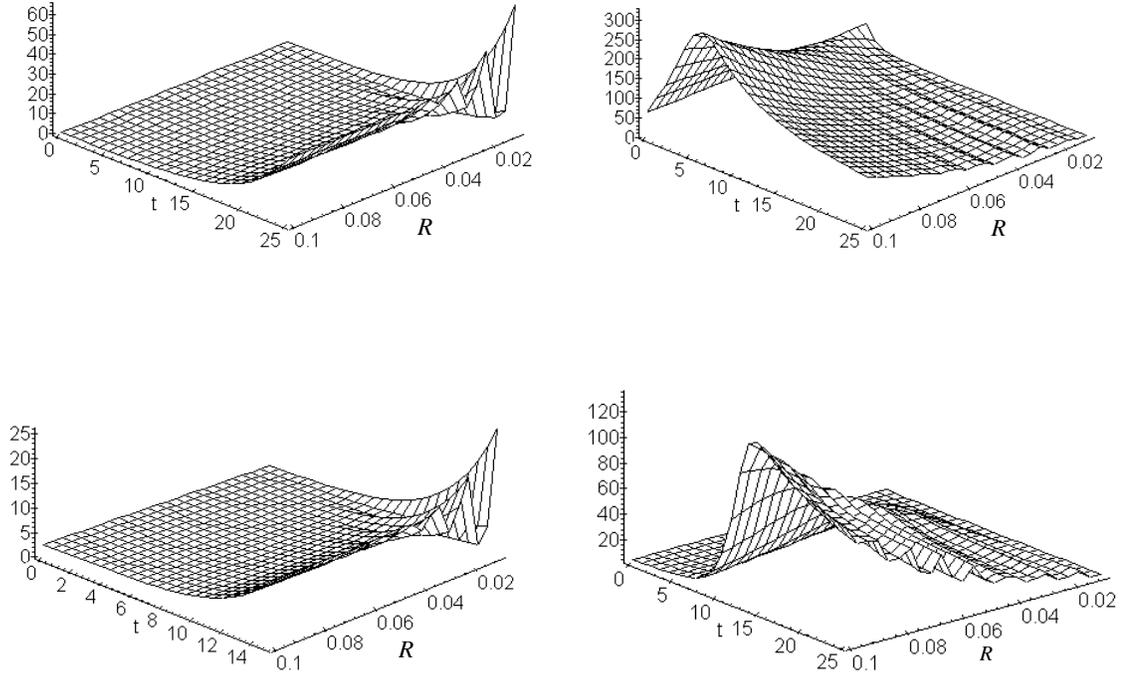

**Fig. 10** Evolution of the amplitude $A(t,R)$ in nm and number of domains $N(t,R)$ for models A and B

The most interesting behavior is demonstrated by $A$ and $N$ as functions of the film radius. Since both quantities are sums of Fourier components, the role of $R$ consists in modulation of the spectra of the steady state waves present on the film surface. Obviously, due to the discrete spectra it is expected to see some resonances reflecting coincidence of a steady state wave with the theoretical most rapid wave with $q_{mq}$. As is seen the amplitude of the waves decreases with increase of the film radius but the dependence is not monotonous one. Such a periodic behavior corresponds well to our expectations for a resonant dependence on $R$. The other two plots presenting the evolution of $N$ demonstrate also a resonant effect. Additionally an increase on the number of uncorrelated domains is observed, which is in accordance with the idea for a finite correlation length of the waves. The latter insists a linear growth of $N$ on the area of the film, which is easily recognized on the figures above. A quantitative measure of the periodicity on $R$ could be made as follows. It is known that the zeros of the Bessel functions are almost regularly distributed and the mean distance between them is about 3. On the other hand the most important wave vector for the evolution of the surface waves is the most rapid $q_{mq}$. Hence, one can propose the following estimate for the period on minima and maxima shown in Figure 10, $\delta R = 3/q_{mq} \approx 0.01$ mm. In the calculation of the latter the previous estimates of $q_{mq}$ are employed, which seem to be similar for both models A and B. Comparing this result with the calculations above confirms our conclusion that the periodicity of $A$ and $N$ on $R$ is a result from the discrete nature of the steady state wave spectrum. Of course, the imposed by us boundary conditions on the film rim are quite idealized. For this reason the effect described above could be strongly depressed in the real TLFs.